# Epitaxial Growth and Electronic Properties of Quasi–Free-Standing Rhombohedral WSe$_2$ Bilayers on Cubic W(110)

Niels Chapuis[1], Meryem Bouaziz[1], Eva Desgué[2], Iann Gerber[3], François Bertarn[4], Pierre Legagneux[2], Fabrice Oehler[1], Julien Chaste[1], Abdelkarim Ouerghi[1*]

[1]*Université Paris-Saclay, CNRS, Centre de Nanosciences et de Nanotechnologies, 91120, Palaiseau, Paris, France*
[2]*Thales Research & Technology (TRT), 91767 Palaiseau, France*
[3]*Université de Toulouse, INSA-CNRS-UPS, LPCNO, 135 Avenue de Rangueil, 31077 Toulouse, France*
[4]*Synchrotron SOLEIL, L'Orme des Merisiers, Départementale 128, 91190 Saint-Aubin, France*

**ABSTRACT.**

Rhombohedral-stacked transition metal dichalcogenides (TMDs) break inversion symmetry between adjacent layers, giving rise to an intrinsic out-of-plane ferroelectric polarization. Controlling the formation of this stacking polytype is therefore essential for harnessing ferroelectric effects in two-dimensional materials. In this work, we demonstrate the epitaxial growth of rhombohedral bilayer tungsten diselenide (3R-WSe$_2$) on a cubic W(110) single crystal by molecular beam epitaxy. We show that selenium passivation of the substrate is key to enable a quasi–van der Waals epitaxy, effectively suppressing strong interfacial bonding and promoting the growth of quasi free-standing bilayer films. The 3R stacking order is confirmed through a combination of Raman spectroscopy and high-resolution angle-resolved photoemission spectroscopy (ARPES), supported by density functional theory (DFT) calculations. ARPES and DFT reveals an indirect-gap electronic structure with the valence-band maximum at the Γ point, as well as a pronounced spin–orbit-driven splitting of 520 ± 20 meV at the K point. Analysis of the measured dispersions yields hole effective masses of 0.46 ± 0.04 m$_e$ and 0.75 ± 0.06 m$_e$ for the upper and lower valence bands at K point, respectively. These results establish a robust route for synthesizing quasi–free-standing 3R-WSe$_2$ and provide a platform for exploring the electronic, optical, and ferroelectric functionalities that emerge from inversion-symmetry breaking in layered TMDs. Our findings further highlight the potential of cubic substrates for deterministic fabrication of rhombohedral TMD heterostructures and ferroelectric devices at the nanoscale.

*Corresponding author: abdelkarim.ouerghi@c2n.upsaclay.fr

## I. INTRODUCTION

Two-dimensional (2D) semiconducting materials exhibit high potential for the next-generation integrated electronic and optoelectronic systems, due to their remarkable physical and structural properties [1]. Among 2D compounds, the transition metal dichalcogenides (TMDs) family with the formula $MX_2$ (M= transition metal, X =Se, S, Te) is of peculiar interest because of its unique properties in the few-layer limit, i.e. large exciton binding energy, wide spin splitting of the valence band and well-defined valley degrees of freedom [2]. One of its promising feature is the indirect-to-direct bandgap transition upon decreasing thickness from bulk to few- or monolayer (1ML) [3,4]. The bilayer (2ML) regime provides even richer physics with the creation of various moiré reconstructions superlattices induced by the in-plane rotation angle between the two layers, which modulates the band structure [5–7]. In their most symmetric multilayer form, TMDs exhibit a variety of polytypes with either 0° or 60° in-plane rotation angle [8], which in the bilayer regime can be summed up to the hexagonal 2H (AA'-type) or the rhombohedral 3R (AB-type) arrangement [9]. The latter exhibits a broken inversion symmetry [10,11], which induces a lift of degeneracies in the electronic structure [9,12], while favouring a permanent ferroelectric polarization [11,13,14].

So far, Chemical Vapor Deposition (CVD) has been one of the most promising methods to synthesize TMDs films, notably with the recent demonstrations of controlled 2H and 3R $WSe_2$ phases growth [15]. Although the high temperature CVD technique has led to large individual mono-crystals of equivalent quality to exfoliated materials [16], large-area growth with controlled in-plane orientation remains challenging. Besides, the control of polytypism by CVD remains a strong challenge, as several multilayer polytypes often coexist on the same substrate [17]. On the opposite, the molecular beam epitaxy (MBE) technique has recently demonstrated significant progresses in the direct growth of large-scale TMDs thin films on various crystalline substrates [18–21] with great control in polytypism [15,22–24].

Due to its reduced symmetry, the stable growth of 3R polytype of few-layer 2D materials is particularly interesting. For graphene, rhombohedrally-stacked multilayer graphene was successfully grown on cubic 3C-SiC(111) substrate [25–27]. Similarly, the direct growth of rhombohedral boron nitride (rBN) growth was reported using particular Ni(100) terraces and Ni(110) bevel facets [28]. For TMD, low-temperature MBE growth can favour 3R growth on Se-treated III-V substrates, but the overall phase purity is challenging as 1T' domains can coexist with the main 3R phase [15]. We propose here an alternative route, based on the idea that square or rectangular surfaces can favour the epitaxial growth of 3R-TMD bilayer without parasitic inclusion. For example, the dense (110) plane of the body-centered cubic (BCC) tungsten single crystal is an interesting candidate for MBE growth of 3R TMD. The use of alternative substrate could also widen the 3R TMD thermal stability window, over the reported temperature-induced 3R-2H phase transition reported for 3R TMD grown on Se-treated III-V substrates [22].

Metal substrates can also serve as ohmic contacts between semiconducting TMDs. In the literature, the reported growth of monolayer $MoS_2$ and $MoSe_2$ over single Au(111) surface evidences the formation of periodic moiré superstructures due to lattice mismatch, which induces local strain in the TMD [29]. Moreover, the lower sub-lattice of chalcogen (S or Se) atoms also strongly interact with the gold surface, resulting in a reduction of the bandgaps in Au supported TMDs compared to $MoS_2$/graphene and $MoSe_2$/graphene counterparts [29–31]. These results suggest that the direct growth on native metal surfaces may not be suitable to obtain unaltered semiconducting TMDs. Instead, specific surface reconstruction using chalcogen termination may be necessary to mimic the van der Waals passivation of native 2D materials [32,33]. Here, we show that introducing a selenization step prior to WSe$_2$ deposition effectively passivates W dangling bonds on the W(110) surface, enabling a quasi–van der Waals interface and markedly reducing film–substrate interactions. We investigate the resulting bilayer WSe$_2$ synthesized by MBE on Se-terminated W(110) using a combined suite of structural and spectroscopic techniques. Low-energy electron diffraction (LEED) and atomic force microscopy (AFM) highlight a high-quality epitaxial growth, while micro-Raman spectroscopy, ARPES and DFT calculations confirm that the film adopts the rhombohedral 3R stacking order. This work provides a pathway toward scalable, substrate-controlled synthesis of large-area rhombohedral TMD bilayers, opening new opportunities for engineering polarization, electronic, and valleytronic functionalities in inversion-symmetry-broken 2D materials.

## II. METHODS

1-cm W(110) substrates were purchased from Surface preparation Laboratory (SPL), Wormerveer. The surfaces were firstly outgassed at 950°C during 60 min in the buffer chamber ($\sim 3 \times 10^{-10}$ mbar) before being transferred to a 2-inch MBE reactor supplied by Dr Eberl MBE-Komponenten dedicated to TMD growth fitted with a valved selenium cracker cell and an electron gun as tungsten source with high purity (4N) Tungsten. The selenium tank filled with ultra-high purity (7N) Se is heated to 290°C, while the cracker temperature is maintained at 600°C. Thorough this study, the tungsten growth rate of 0.002 Å.s$^{-1}$ as well as the selenium growth rate of 0.5 Å.s$^{-1}$ were calibrated by a quartz crystal monitor. We monitor the growth thickness through a second quartz crystal positioned next to the electron gun. The growth temperature is set at 400°C. In order to favour the crystal quality of the layer, we interrupt the growth after each nominal ½ ML and anneal the surface under Se at 400°C for 5 minutes. At the end of the growth process, a subsequent annealing step under Se at 600°C is applied. The surface crystallography is monitored during growth by *in operando* RHEED, whereas complementary analysis is performed using *in situ* LEED at 48 eV. In order to protect the epi-layer from external reaction, the samples are capped with a 50 nm amorphous selenium layer.

Raman spectroscopy measurements were performed on a Horiba Scientific LabRAM HR instrument at an excitation of λ = 532 nm, in a backscattering geometry in a parallel-polarized configuration with a 360° rotational sample stage. The spectral resolution was ~0.7 cm$^{-1}$ for a grating with 1800 grooves per mm. The spectrometer was calibrated with a pristine silicon sample.

ARPES experiments were performed at the CASSIOPEE beamline of the SOLEIL synchrotron light source. The CASSIOPEE beamline is equipped with a Scienta R4000 hemispherical electron analyser whose angular acceptance is ±15° (Scienta Wide Angle Lens). The experiment was performed at T = 30K. The total angle and energy resolutions were 0.25° and 16 meV, respectively.

All calculations were performed using the VASP package [34] within the projector augmented-wave (PAW) formalism [35,36], treating 14 and 6 electrons explicitly as valence states for W and Se, respectively. The first Brillouin zone was sampled using a Γ-centered (12 × 12 × 1) k-point grid with a Gaussian smearing of 0.05 eV to account for partial band occupancies. Structural optimizations were carried out using the PBE-D3 functional [37] to accurately describe the weak van der Waals interactions between layers. The in-plane lattice constant was fixed at 3.32 Å, and a vacuum of ~22 Å was applied along the z-direction to prevent spurious interactions between periodic images. On top of these calculations, HSE06 hybrid functional [38–40] calculations including spin-orbit coupling were performed. Wannier-interpolated band structures were obtained using the WANNIER90 package [41].

### III. RESULTS AND DISCUSSION

We start our WSe$_2$ synthesis on commercial W(110) substrates (from SPL. Wormerveer, Netherlands), which are thermally outgassed in ultra-high vacuum (UHV) at 950°C in a dedicated UHV buffer chamber during 60 minutes. The surfaces exhibit a clear $(1 \times 1)$ LEED reconstruction (Figure 1(a)) associated to the bare and oxide-free W(110) surface. This clean W(110) surface is first exposed to a selenium flux (0.5 Å/s), which leads to an obvious change in the LEED pattern, close to a $(1 \times 3)$ reconstruction (Figure 1(b)) for all surfaces temperatures exceeding 500°C (see Figure S1 (a-f) within the Supplemental Material, SM). We then select an optimal annealing temperature of 600°C. The observed LEED pattern evolution is indicative of a sub-monolayer adsorption of Se on W(110), with a saturation coverage of about 0.6 ML [42], and similar evolutions have been reported for the adsorption of indium [43] and antimony [44] on W(110). Besides, this Se pre-exposure also resemble to the Se-passivation of GaAs(111)$_B$ [20] or GaP(111)$_B$ [45] substrates used for quasi-van der Waals epitaxy. The detailed procedure of the WSe$_2$ bilayer growth is described in the **Methods**. In brief, we set a growth temperature of 400°C, followed by a subsequent annealing step at 600°C under Se. As reported in previous studies, the 1T' metastable polymorph is not forming at growth temperatures exceeding 400°C [46], while the 3R polytype should undergo a permanent phase transition into 2H under Se annealing above 400°C. In Figure 1(c), the LEED pattern obtained after a sub-monolayer WSe$_2$ growth exhibits both the Se-terminated W(110) $(1 \times 3)$ pattern an additional $(1 \times 1)$ hexagonal contribution, which we attribute to epitaxial WSe$_2$. We observe that one of axes of the hexagonal $(1 \times 1)$ aligns in reciprocal space with the [1-10] direction of W(110). So that the final alignment in real space is $[1\bar{1}00]_{WSe_2}/[001]_{W(110)}$ and $[\bar{1}\bar{1}20]_{WSe_2}/[1\bar{1}0]_{W(110)}$ in plane, leading to $[0001]_{WSe_2}/[110]_{W(110)}$ out-of-plane, as schematized in Figure 1(e). We note that the W lattice is symmetrical by inversion, whereas the WSe$_2$ orthorhombic lattice is not, which could lead to the formation of 180° twins with two possible orientations. In Figure 1(d), the $(1 \times 1)$ hexagonal pattern resulting from a bilayer growth confirms the formation of epitaxial WSe$_2$. After calibrating the LEED with W(110) spots of lattice parameter 3.16 Å [47], we deduce a resulting in-plane lattice periodicity of 3.4 Å, slightly higher than the theoretical value for WSe$_2$ (3.30 Å [48]). It highlights the preferential orientation of the domains along the primary directions of the hexagonal lattice with a spread angle of about 6°. Besides, the resulting topography measured with AFM (see Figure S(2) within the SM) highlights a rather smooth surface (RMS~0.8 nm).

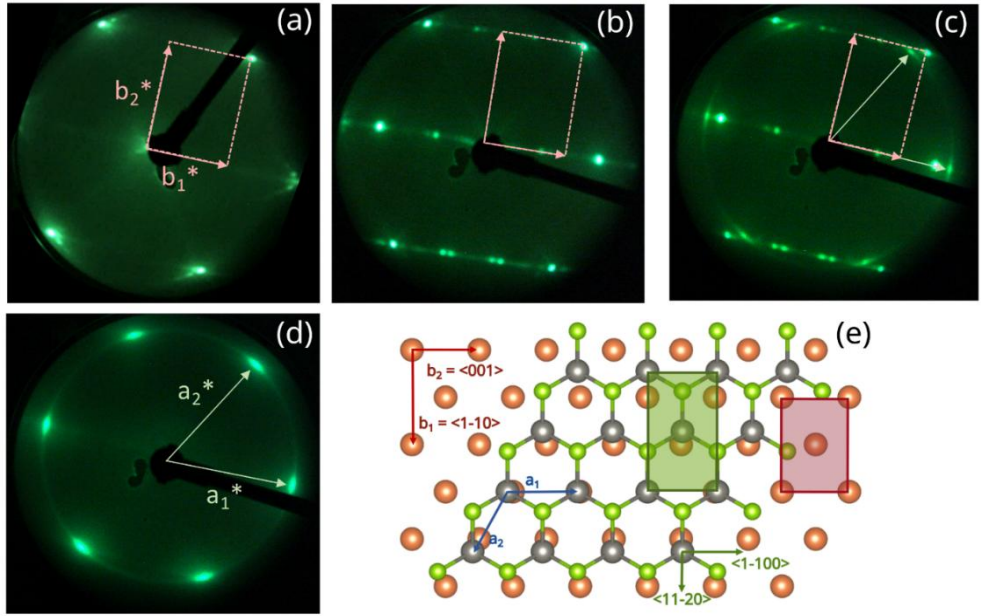

**Figure 1.** LEED pattern recorded at 48 eV of an outgassed W(110) surface (a), a Se-terminated substrate (b), a sub-monolayer WSe$_2$ (c), a grown WSe$_2$ bilayer (d) and the associated epitaxial relationship schematic drawn with VESTA (e). The W atoms from the substrates are tinted in orange, and the W(110) and orthorhombic WSe$_2$ lattices are highlighted in red and green, respectively.

The associated Raman spectra obtained at five different zones over the sample (Figure 2(a)) evidence the homogeneity of the film at the macroscopic scale, with similar signals composed of four different contributions. The experimental spectrum (black-dot curve) exhibits two dominant contributions at 251.5 cm$^{-1}$ and 248.4 cm$^{-1}$, respectfully attributed to the WSe$_2$ A$_{1g}$ and E'$_{2g}$ modes [49], while the weaker feature at 259 cm$^{-1}$ is associated to the acoustic 2LA(M) mode [50]. This combination of Raman data and LEED pattern demonstrates the existence of the epitaxial WSe$_2$ over W(110) with crystal unit cell close to relaxed bulk WSe$_2$. In the following, we now investigate the particular polytype of WSe$_2$. The residual contribution at 310.5 cm$^{-1}$ (Figure 2(a)) is generally attributed to the B'$_{2g}$ vibration mode, which results from the van der Waals force interaction between several adjacent layers. This is coherent with our expected 2ML thickness from previous growth calibration. In Figure 2(b), we present the decomposition of the Raman spectrum as well as reference spectra of 3R and 2H stacked bilayer WSe$_2$ obtained by CVD. The comparison between our MBE-grown WSe$_2$ sample (black-dot curve) and the CVD-grown 2H and 3R WSe$_2$ references (cyan and yellow curves, respectively) for which the 3R or 2H character has been unequivocally determined [15,51] allows us to confirm the 3R character of our bilayer. In CVD-grown samples, we observe that the A$_{1g}$ mode is strongly dominant over E'$_{2g}$ in the case of the 2H WSe$_2$ bilayer, while the 3R configuration induces the notable reduction of A$_{1g}$, leading to the observation of both vibration modes with equivalent intensities [50]. The matching position of the two dominant contributions between the CVD 3R reference and our MBE-grown sample indicates on some 3R character, as already observed in MBE grown 3R WSe$_2$/GaP(111) heterostructure [22,24]. The line width of the A$_{1g}$ and E'$_{2g}$ contributions are 2.7 cm$^{-1}$ and 6.2 cm$^{-1}$, respectfully, calculated by the full width at half-maximum (fwhm) after a Lorentzian fitting, which indicates a larger defect density than the CVD 3R reference (A$_{1g}$~1.3 cm$^{-1}$; E'$_{2g}$~2.2 cm$^{-1}$ fwhm). Additional Raman mapping (Figure 2(c-d)) confirms the homogeneity of the A$_{1g}$ and E'$_{2g}$ signals at the macroscopic scale, with enhanced intensities along intrinsic grain boundaries of the substrate, yet without any significant change in position or

$A_{1g}$/$E'_{2g}$ intensity ratio (see Figure S3(b-d) within the SM). Lastly, the position variation between the 2LA(M) and $A_{1g}$ contributions (~6.7 cm$^{-1}$) as well as the $A_{1g}$/2LA(M) intensity ratio over the selected area (see Figure S3(e-f) within the SM), close to the CVD 3R reference, tend to confirm the homogeneous formation of the 3R phase.

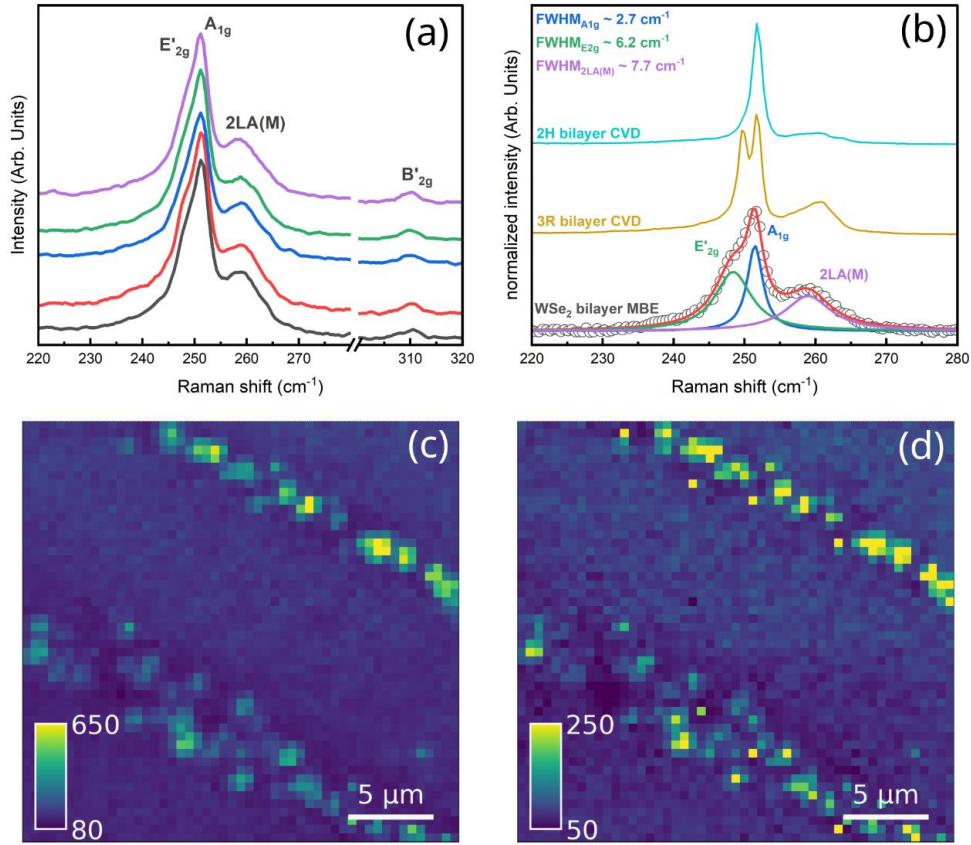

**Figure 2.** Raman spectra measured at five different positions on the sample (a). Normalized Raman spectrum acquired at zone 2, together with its peak decomposition and comparison with CVD-grown bilayer 3R-WSe$_2$ and 2H-WSe$_2$ reference spectra (b). Raman intensity maps (25 × 25 µm²) of the $A_{1g}$ (c) and $E'_{2g}$ (d) modes, illustrating the spatial uniformity of the film (the two linear features arise from substrate scratches).

While the previous micro-Raman and LEED data indicate that our WSe$_2$ grown on W(110) is a van der Waals hybrid 2D/3D heterostructure, we still need to prove the absence of covalent bonds between the different layers within our sample. To achieve this, we use X-ray photoemission spectroscopy (XPS) acquired at synchrotron facility in order to study the core levels of the different atoms [52]. XPS overview of the bilayer WSe$_2$ evidences sharp peaks corresponding to the W 4f, Se 3d and the valence band (Figure 3(a)). The binding energy plot of both Se 3d and W 4f core levels (CLs) are presented in Figure 3(b-c). The W 4f CL is decomposed into a first spin-orbit doublet W 4f$_{7/2}$ and W4f$_{5/2}$ at binding energy (BE) 32.7 and 34.9 eV, and a second doublet at BE W 4f$_{7/2}$=31.9 eV and W 4f$_{5/2}$=34.1 eV. The Se 3d CL is decomposed into a single spin-orbit doublet Se 3d$_{5/2}$ and Se 3d$_{3/2}$ at BE 54.9 and 55.8 eV, respectively. In both spectra, the most intense doublet is associated to the signature of the WSe$_2$ layer, while the second lower component in W 4f CL is attributed to the tungsten substrate. We link the additional weak doublet observed in the W 4f CL captured with hν = 80 eV at BE W 4f$_{7/2}$=32.1 eV and W 4f$_{5/2}$=34.3 eV to the Se-terminated interface (Figure 3(d)), which displays a quasi-van der Waals character. No signature of components at other binding energies that could be attributed to interfacial reactions is observed, confirming that the grown WSe₂ crystal is obtained by quasi–van der Waals epitaxy. Consistently, the LEED patterns of WSe₂ displays a well-ordered surface with an angular spread of approximately ±3° and no detectable moiré periodicity.

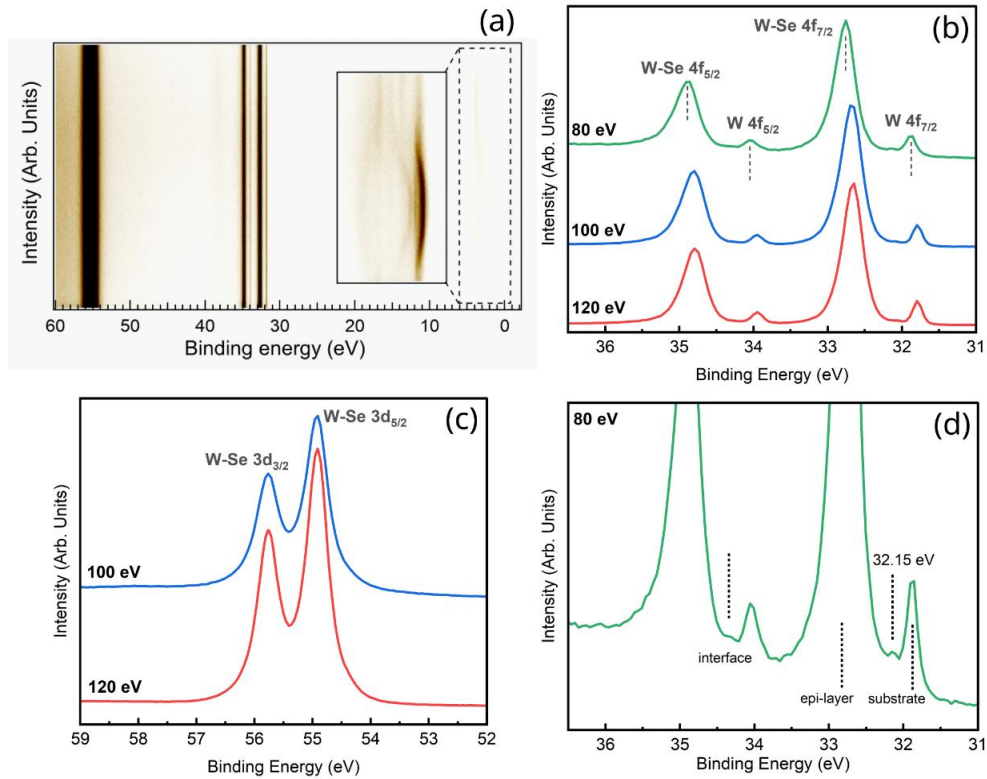

**Figure 3.** XPS spectrum measured with hν = 80 eV, the mapping represents the recorded intensity along the binding energy range (a). XPS core level spectra of W 4f (b) and Se 3d (c) recorded with hν = 80-100-120 eV. XPS core level spectra of W 4f recorded with hν = 80 eV (d).

To investigate the electronic properties of the WSe₂/W(110) heterostructure, we performed ARPES measurements at 30 K, complemented by density functional theory (DFT) calculations. Isoenergy contours at $E-E_F=2.25$ eV (Figure 4(a)) reveal well-defined hexagonal pockets, indicating that the WSe₂ film forms a single in-plane crystallographic domain with no evidence of rotational domain mixing within the beam spot (50−μm). Figure 4(b) shows the ARPES intensity map along the Γ–K–M high-symmetry direction using a photon energy of 60 eV. The sharpness of the bands at the G and K point reflects the high quality of the sample. Along the high symmetry ΓK direction, we resolve two split branches of the valence band of WSe₂ at the K point. Based on the measured energy distribution curves (EDC) at K, we estimate that the spin-orbit coupling (SOC) is about 520 ± 20 meV. The large experimental splitting of the valence band at K is one of the signatures of the strong atomic spin–orbit interaction in this compound. The valence band maximum (VBM) is clearly located at Γ, mostly composed of W $d_{z^2}$ orbitals and $p_z$ orbitals of Se. At the K point, it is formed by the $d(xy)$ and $d(x^2-y^2)$ orbitals of tungsten, hybridized with $3p(x+y)$ of the chalcogens, as it results from DFT calculations.

To enhance fine spectral features and clarify the band structure shown in Figure 4(b), second derivatives of the photoelectron intensity with respect to energy and momentum are presented in Figure 4(d). The measured band structure was compared to our HSE-based DFT calculations including spin-orbit coupling for 3R-stacked bilayer WSe₂ (Figure 4(c)). All main features are well reproduced. Both the shape and energetic positions of the valence bands near Γ and K match the predictions for 3R stacking. In particular, the lack of inversion symmetry in the 3R structure gives rise to pronounced spin splitting at the K valleys and shifts the VBM from K toward Γ. The dispersion and spectral weight of the deeper valence bands also agree with the 3R DFT model, confirming that bilayer WSe$_2$ grown on W(110) crystallizes in the rhombohedral stacking sequence. These results also evidence that a rectangular substrate surface favors the epitaxial growth of 3R-TMD bilayer, in good agreement with the recent successful growth of rhombohedral-stacked graphene on 3C-SiC(111) [25–27] and rBN on Ni(100) terraces [28].

The valence band maximum at Γ/K is experimentally determined by ARPES and confirmed by DFT, showing a near degeneracy (~30 meV). According to DFT, the conduction band minimum lies at the Q point along Γ–K, confirming the indirect nature of the band gap in bilayer WSe$_2$. Furthermore, ARPES allows us to quantify the band curvature and extract the hole effective masses for bilayer WSe$_2$ on W(110). We fitted the experimental dispersion within ±0.1 Å$^{-1}$ around the band extrema using a parabolic function to extract the hole effective masses. At K, we obtain effective masses of 0.46 ± 0.04 m$_e$ for the upper band and 0.75 ± 0.06 m$_e$ for the lower band (where m$_e$ is the free electron mass), highlighting the clear asymmetry between the two branches. These values are consistent with previously reported data and confirm the pronounced asymmetry of the valence band. These experimentally derived values agree with HSE06 hybrid functional calculations, confirming the high crystalline and electronic quality of the epitaxial rhombohedral bilayer. Finally, by comparing our data with DFT calculations for freestanding bilayer WSe$_2$ and with the LEED and XPS image <span style="color:red">(Figure 1(d) & 3(b))</span>, we find that the electronic dispersion of our bilayer WSe$_2$/W(110) hybrid heterostructure closely matches that of a freestanding 3R bilayer WSe$_2$. Considering this agreement, we conclude that the bilayer WSe$_2$ behaves as a quasi-freestanding film on the cubic substrate, exhibiting negligible interlayer hybridization. This behavior contrasts markedly with previous reports on TMDC growth on noble-metal substrates such as Au(111) [30,53,54], where strong substrate interactions and pronounced hybridization are commonly observed. Overall, these findings identify Se-terminated W(110) as a structurally and electronically compatible for the epitaxial growth of transition-metal dichalcogenides. The MBE-grown 3R-WSe$_2$ bilayer is uniform over the full ~7 mm diameter sample, as confirmed by structural characterization. ARPES measurements were carried out at multiple randomly chosen locations with an 80 × 80 µm² beam spot, and identical electronic features were observed at all measured positions, confirming the macroscopic homogeneity of the film.

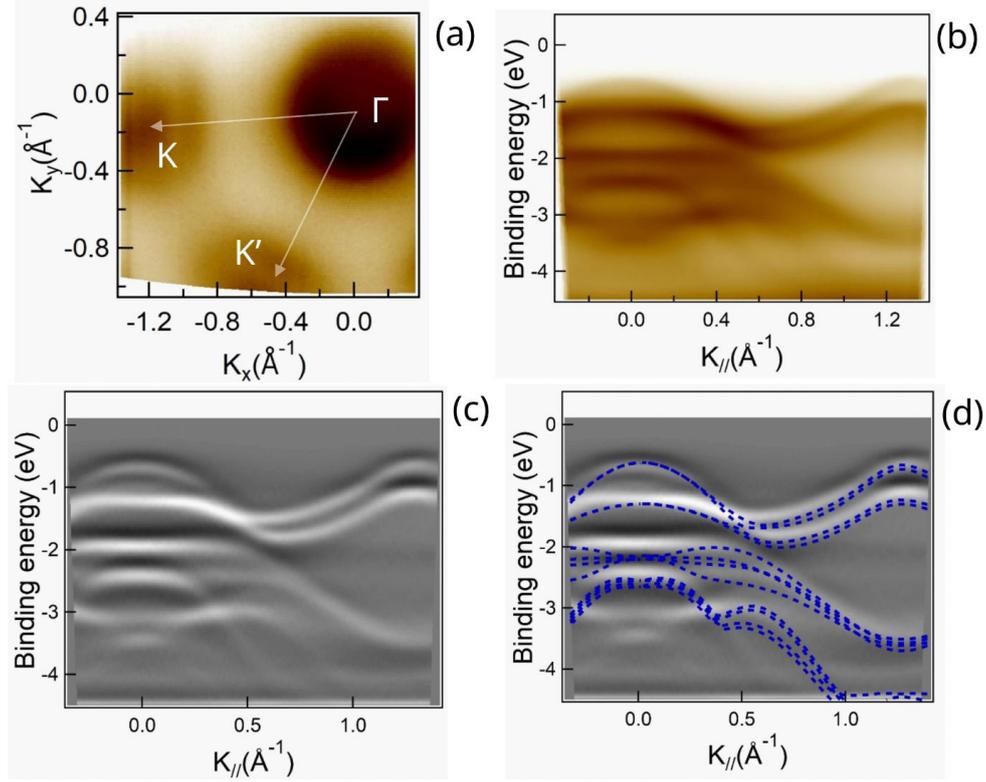

**Figure 4.** ARPES measurement (hν = 60 eV, T = 30 K) of the electronic structure along the ΓKM high-symmetry directions (a), with the corresponding second derivative of the raw scans (b); same data with overlaid theoretical DFT calculations from freestanding bilayer 3R stacking order WSe$_2$ (the Fermi level position is located at zero of the binding energy) (c). Isoenergy cut of the first Brillouin zone of the ΓKM plane (d).

## CONCLUSIONS

In summary, we demonstrate the epitaxial growth of rhombohedral-stacked bilayer WSe$_2$ on a W(110) single crystal via molecular beam epitaxy. Combining micro-Raman, LEED, and XPS measurements, we confirm that the sample consists of epitaxial 3R-WSe$_2$ bilayers. These layers exhibit weak hybridization with the substrate, and their electronic band structure is well described by calculations for freestanding 3R-WSe$_2$. This conclusion is further supported by ARPES measurements and DFT calculations, which reveal an indirect band gap with the VBM located at Γ. We directly observe strong spin splitting near the K point, with values in excellent agreement with DFT calculations. These results establish that cubic substrates provide a suitable platform for van der Waals epitaxy of 3R-stacked TMDs, opening promising avenues for exploiting their electronic and spintronic properties in future nanoelectronic and optoelectronic devices.

**ACKNOWLEDGMENTS:** We thank Chiara Bigi for his support during our experiments at the CASSIOPEE beamline. This work was supported by the French National Research Agency (ANR) through the projects COMODES (ANR-22-CE09-0021), 2DonDemand (ANR-20-CE09-0026), DEEP2D (ANR-22-CE09-0013), MixDferro (ANR-21-CE09-0029), and FastNano (ANR-22-PEXD-0006), as well as by the "Investissements d'Avenir" program (Labex NanoSaclay, ANR-10-LABX-0035) and the RENATECH network. I. C. Gerber acknowledges CALMIP (Project No. p0812) and GENCI-TGCC/IDRIS (Grant No. A012096649) for computational resources.

**Competing financial interests:** There are no conflicts to declare.

**Data availability:** The datasets generated during and/or analyzed during the current study are available from the corresponding author on reasonable request.

**Supplementary S1: LEED characterization of the W(110) surface with the Se-annealing step**

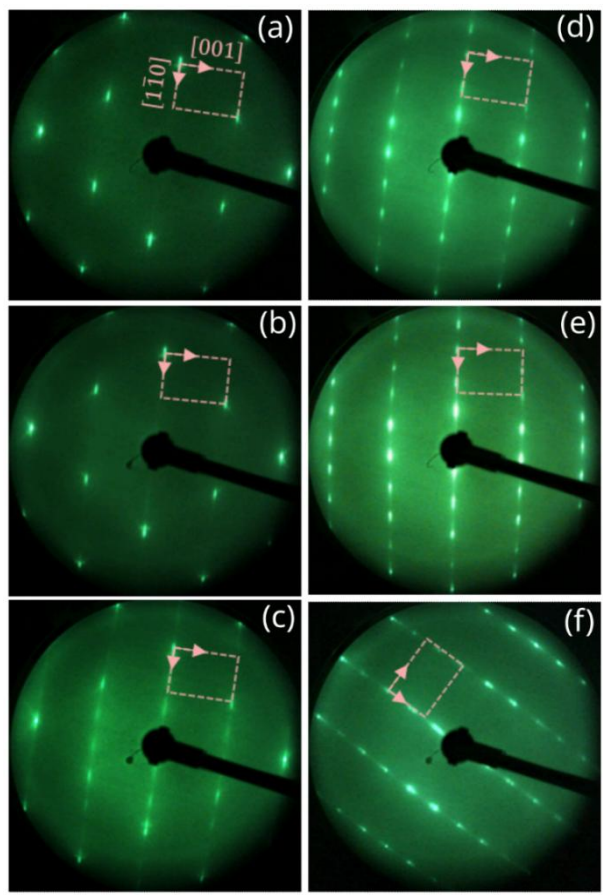

**Figure S1.** Evolution of the LEED pattern at 150 eV of the bare W(110) surface (a) and after a 30min Se exposure at 300°C (b), 400°C (c), 500°C (d), 600°C (e) and 800°C (f).

The evolution of the LEED pattern with the surface selenization temperature evidences a surface reaction above 400°C, with the appearance of blur intermediate dots in Figure S1c. When increasing the temperature up to 500-600°C (Fig. S1d-e), those dots appears clearer and sharper. Since the somewhat (1 × 3) pattern does not evolve between 500°C and 800°C, we assume that the blurriness of the intermediate dots at low temperature originates from the resulting poor crystal quality of the Se-terminated surface.

**Supplementary S2: AFM characterization of the grown WSe$_2$ bilayer**

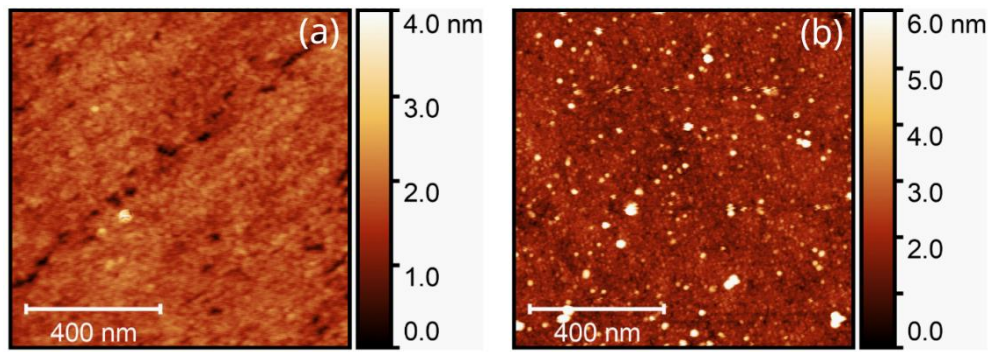

**Figure S2.** $1 \times 1$ µm² AFM image captured on the bare W(110) surface (a) and after a bilayer WSe$_2$ growth (b).

The surface morphology is investigated by *ex situ* AFM on a ParkNX10 model in non-contact mode using standard AC160TS probes with a resonance frequency at ~300 kHz. The RMS of about 0.8 nm is obtained after a proper image analysis on Gwyddion.

**Supplementary S3: Additional Raman mapping**

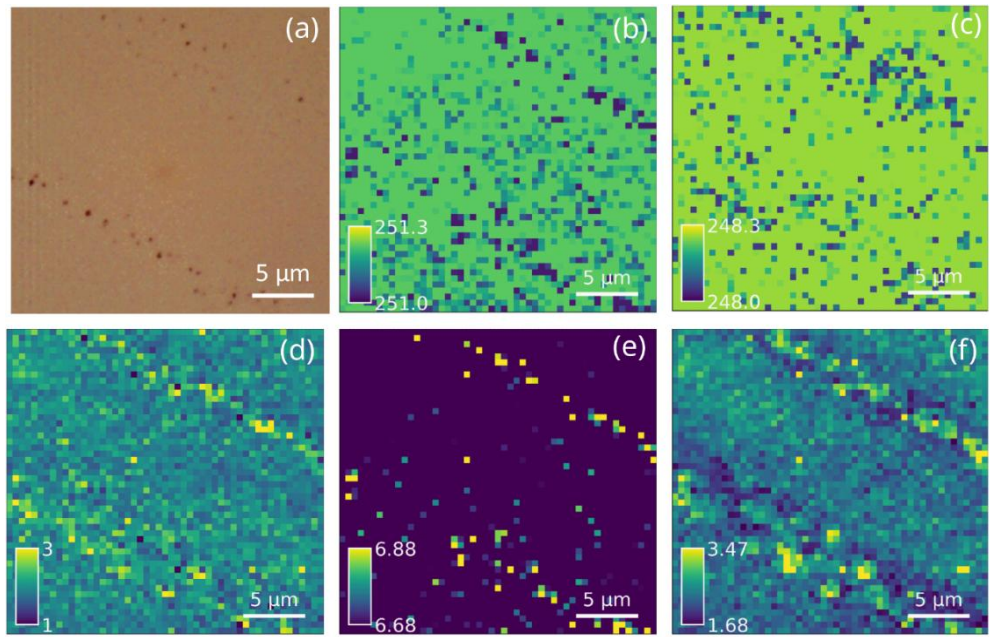

**Figure S3.** $25 \times 25$ µm² optical microscopy image (a) with the associated Raman map of the A$_{1g}$ (b), E'$_{2g}$ (c) position and of the intensity ratio between A$_{1g}$ and E'$_{2g}$ (d). Raman map of the position variation of 2LA(M) with respect to A$_{1g}$ (e), and of the intensity ratio between A$_{1g}$ and 2LA(M) (e).